\documentclass[a4paper, 10 pt, journal]{IEEEtran}
\usepackage{cite}
\usepackage{graphicx}
\usepackage{graphics}
\usepackage{color}
\usepackage{colortbl}
\usepackage{xcolor}
\usepackage[utf8]{inputenc}
\usepackage[english]{babel}
\usepackage{amsmath}

\usepackage{amsthm}
\usepackage{amssymb}  
\usepackage{setspace}

\usepackage{enumitem}
\usepackage{centernot}
\usepackage{bm}
\usepackage{placeins}
\usepackage{hyperref}
\usepackage{cleveref}
\usepackage{picins}
\usepackage{wrapfig}
\usepackage{import}
\usepackage{mcode}
\newtheorem{theorem}{Theorem}
\usepackage{comment}

\IEEEoverridecommandlockouts                              

\begin{document} 

\title{String Stability of Connected Vehicle Platoons under Lossy V2V Communication}
%
%
%

\author{Vamsi~Vegamoor,~\IEEEmembership{IEEE~Student~Member}, Sivakumar~Rathinam,~\IEEEmembership{IEEE~Senior~Member}
        and~Swaroop~Darbha,~\IEEEmembership{IEEE~Fellow}
\thanks{Vamsi Vegamoor is a doctoral candidate in Mechanical Engineering at Texas A\&M University, College Station, TX 77843, USA; e-mail: vvk@tamu.edu}
\thanks{Sivakumar Rathinam and Swaroop Darbha are faculty members in Mechanical Engineering at Texas A\&M University, College Station, TX 77843-3123, USA; e-mail: \{srathinam, dswaroop\}@tamu.edu}

}

\maketitle

\begin{abstract}

\color{black} Recent advances in vehicle connectivity have allowed formation of autonomous vehicle platoons for improved mobility and traffic throughput. In order to avoid a pile-up in such platoons, it is important to ensure platoon (string) stability, which is the focus of this work\color{black}. As per conventional definition of string stability, the power (2-norm) of the spacing error signals should not amplify downstream in a platoon. But in practice, it is the infinity-norm of the spacing error signal that dictates whether a collision occurs. We address this discrepancy in the first part of our work, where we reconsider string stability from a safety perspective and develop an upper limit on the maximum spacing error in a homogeneous platoon as a function of the acceleration maneuver of the lead vehicle. In the second part of this paper, we extend our previous results by providing the minimum achievable time headway for platoons with two-predecessor lookup schemes experiencing burst-noise packet losses. Finally, we utilize throttle and brake maps to develop a longitudinal vehicle model and validate it against a Lincoln MKZ which is then used for numerical corroboration of the proposed time headway selection algorithms.
\end{abstract}

\section{Introduction}\label{sec:Intro}
Vehicle platooning has been studied since the late 1950s, with early efforts on car following \cite{1958Platooning} simply attempting to replicate human driving behaviour. Modern approaches to platooning focus on achieving small inter-vehicle spacing to improve traffic throughput (mobility) as well as reduce fuel consumption \cite{ThroughputPaper,fuel1}. At the same time, a sufficient headway has to be maintained between vehicles to prevent pile-ups. Consequently, a majority of contemporary research on vehicle platooning involves achieving the smallest possible inter-vehicle spacing while guaranteeing safety. 

\begin{figure}[htbp]
	\centering
	\includegraphics[width=0.45\textwidth]{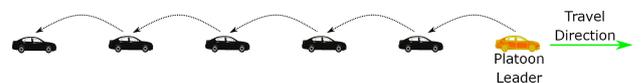}
	\caption{{CACC} platoon using information from one immediately preceding vehicle.}
	\label{fig:1VehAVFS}
\end{figure}

\begin{figure}[htbp]
	\centering
	\includegraphics[width=0.45\textwidth]{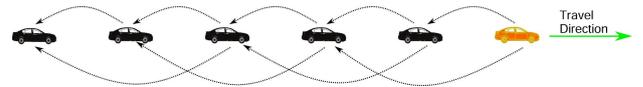}
	\caption{{CACC+} platoon using information from two preceding vehicles.}
	\label{fig:2VehAVFS}
\end{figure}

Adaptive Cruise Control (ACC) systems are now widely available on passenger vehicles. These use onboard sensors (typically radar) to measure the relative velocity and distance to the preceding vehicle. This information is then used in a servomechanism to supply throttle or brake input to the ego vehicle. Cooperative Adaptive Cruise Control (CACC) systems utilize additional information (typically acceleration) transmitted directly from the preceding vehicle using wireless Vehicle-to-Vehicle (V2V) communication. Advanced cooperative systems implement more complex communication typologies and can utilize information from multiple preceding or succeeding vehicles. In this work, we focus on one-vehicle (CACC) and two-vehicle (CACC+) predecessor lookup schemes. \textcolor{black}{For the purpose of this paper, ACC refers to the control scheme that is restricted to using on-board information, CACC refers to a scheme that utilizes acceleration information from the immediate vehicle ahead and CACC+ refers to a platoon system that utilizes acceleration measurements from both the vehicles ahead. For the latter two schemes, the information is communicated wirelessly. \color{black} For clarity, Figs. \ref{fig:1VehAVFS} and \ref{fig:2VehAVFS} provide a visualization of the information flow for CACC and CACC+ systems.}

\color{black}
For a platoon to be string stable, any fluctuations that occur at the head of a platoon need to be damped out as they propagate towards the tail. This condition is often expressed in the frequency domain \cite{swaroop1994phd} which ensures that the 2-norm of spacing errors (i.e., the power of spacing error signals) do not amplify. However from a safety perspective, it is the maximum (infinity norm) of the spacing error signal that dictates if a collision will occur. \textcolor{black}{A further discussion of this issue is presented in \cite{vegamoorAVFSReview}}. In the current work, we develop an upper bound for the maximum spacing error as a function of the acceleration maneuver of the lead vehicle. This result, presented for both CACC\footnote{A portion of this paper has been accepted for publication at the IEEE ITSC 2020 conference \cite{vegamoor2020mobility}; a pre-print of the conference paper is \href{https://arxiv.org/abs/2003.04511}{ available online}.}  \cite{vegamoor2020mobility} and CACC+ schemes, can also be used to pick a safe standstill distance, \color{black}which many previous works either ignore (set to zero) or pick arbitrarily. Ignoring the standstill distance is not practical as it would mean that the bumpers of vehicles touch when the vehicles are stationary. At the same time, arbitrarily choosing this parameter risks an unnecessary increase in inter-vehicular spacing which needs to be minimized as much as possible, for improved fuel efficiency and mobility.

\color{black}
To analyze string stability, all vehicles in the platoon are modeled as point masses with a parasitic lag $\tau$:
\begin{equation}\label{eqn:PointMassModel}
    \ddot{x}_i=a_i, \quad \tau \dot{a}_i+a_i=u_i,
\end{equation}
where $x_i$ is the position of the $i^{th}$ vehicle, $a_i$ its acceleration and $u_i$ is the control input. To account for heterogeneity in individual parasitic lags, $\tau$ may be chosen as the maximum value in the platoon. Previous research from California PATH \cite{CalPathExample,CalPathExample2} have successfully used such models. As further corroboration for the validity of the linearized approximation of the longitudinal vehicle dynamics, we also present numerical simulations with a higher fidelity model-in-loop (MIL) setup that has been validated against data from a 2017 Lincoln MKZ Hybrid.

It has been long established \cite{swaroop1994phd} that for an ACC platoon, string stability can be guaranteed if the time headway chosen is at least twice the sum of parasitic lags in the vehicle, assuming homogeneity in vehicle capabilities. 
It has also been demonstrated that the time headway can be safely reduced further in CACC platoons using V2V communication\cite{V2VBenefits}. Majority of early work on vehicle platooning ignored imperfections in the V2V links. In reality, wireless channels are prone to packet drops due to interference and/or bandwidth restrictions. In the last decade, researchers have noted that lossy communication channels degrade string stability \cite{LeiITSconf2011,LossyCommsVargas}. Workarounds have also been proposed \cite{GracefulCACC,AccianiObserver_Conf} that estimate the information lost due to dropped packets. The algorithm in \cite{GracefulCACC} does not account for the packet loss rate and consequently enforces a significant penalty on the time headway even if only a few packets are dropped. While \cite{AccianiObserver_Conf} suggests that the minimum stabilizing time headway increases with packet loss rates, it does not provide an express relationship between headway and the effective loss rate. Moreover, both of them enforce additional computational burden as they require implementing an estimator.

\color{black}
We should note that other researchers have focused on delays in the platoon  instead of lags \cite{DelayPaperPloeg1,DelayPaper2Hedrick,DelayPaper3Besselink}. Zeng et al \cite{JointCommDelay} have also studied stochastic communication delays in platoons and derived bounds on the maximum allowable delay for string stability. That said, the latest test reports from the $5G$ Automotive Association have concluded that the end-to-end latency for platooning applications remained under $100$ \textcolor{black}{ms} \cite{5GAA_Report} in spite of heavy congestion (over $500$ vehicles contributing to interference). On the other hand, we measured the lag in deceleration of a $2017$ Lincoln MKZ to be over $370$ \textcolor{black}{ms} and our own testing revealed communication latency in a non-congested scenario to be under $5$ \textcolor{black}{ms}. Consequently, we have assumed the vehicle model to be dominated by actuation lags ($\tau$). Moreover, packet loss phenomenon is more common \cite{NLOS_UMich} due to loss of Line-Of-Sight in V2V applications, so it is more likely that a packet will be lost completely rather than arrive with a delay.

Finally, our approach also has an advantage over works like \cite{LossyCommsVargas}, \cite{JointCommDelay} which appear to incorrectly conclude that if network performance is poor, string stability cannot be achieved. Our approach allows smooth (and reversible) transition from CACC+, to CACC and then to ACC (no communication) depending on the packet reception performance.

\color{black}

Earlier work from the authors \cite{vegamoor2020mobility, vegamoor2019} proposed a new limit on the minimum time headway for lossy CACC platoons, given a packet reception probability. 
\color{black} Specifically, in the earlier work \cite{vegamoor2019}, we first explored the feasibility of a smooth inverse relationship between packet reception rate and minimum achievable time headway with a simple packet loss model (independent, identically distributed). This work confirmed that a CACC system with packet losses can smoothly transition to ACC depending on the packet reception rate. That is, at every packet reception rate in $[0,1]$, a string stable time headway can be calculated.
Next, in \cite{vegamoor2020mobility}, we relaxed the i.i.d assumption and demonstrated that the same limit still applies even for burst-noise channels, which are accepted to be more representative of wireless networks \cite{PacketLossRev2}.
\color{black}
In this paper, which is an extension of \cite{vegamoor2020mobility}, we discuss the difficulties in obtaining a similar limit for CACC+ (two-vehicle lookup)schemes and instead propose a usable approximation. \textcolor{black}{We also note that the same approximation approach can be used for three or more vehicle lookup schemes, with one caveat, as explained later.}

The three main contribution of this paper are as follows:
\begin{itemize}
    \item Develop a usable bound on the maximum spacing error in a homogeneous platoon, which is more germane to safety and collision prevention compared to traditional criterion for string stability.
    \item Provide a sufficient condition on the minimum string stable time headway for platoons with lossy V2V communication.
    \item Demonstrate the validity of the proposed lossy vehicle follower systems through high fidelity numerical simulations.
\end{itemize}

It should be noted that the results pertaining to CACC+ in this paper are novel, while those for CACC have recently been accepted for IEEE ITS conference presentation and have been replicated here for completeness.

\section{Maximum Spacing Errors in a String}\label{sec:SpacingBound}
Let a platoon consist of $N+1$ vehicles, with the lead vehicle indexed starting from $0$ and each of the automated follower vehicles indexed $i=1,2, \cdots N$. We will use ${\mathcal I}_N:=\{1,2, \ldots, N\}$ to represent the set of follower vehicles. Let the state of the $i^{th}$ vehicle be denoted by \textcolor{black}{ $\zeta_i(t) \in \mathbb{R}^n$}, while $y_i(t)$ represents its output (such as spacing errors or its derivatives with respect to some origin). Let the disturbance acting on the $i^{th}$ vehicle be $d_i(t)$. We will use ${\cal S}_i$ to denote the set of vehicles whose information is available to the $i^{th}$ vehicle for feedback. In applications such as Adaptive Cruise Control (ACC) and Cooperative Adaptive Cruise Control (CACC), the set of vehicles from which information is available is ${\cal S}_i = \{i-1\}$ for a single preceding vehicle lookup scheme. This scenario is explored in Theorem \ref{thm:outputBound}. Theorem \ref{thm:outputBound2Veh} explores multiple vehicle lookup schemes, where we could have ${\cal S}_i = \{i-1, i-2, \ldots, i-r\},$ where $r$ depends on the connectivity. For some appropriate functions $f_{ij}$ and $h_i$, the evolution of spacing errors may be described by a set of equations of the form:
\begin{eqnarray} \label{eqn:ZetaEOM}
{\dot \zeta}_i = \sum_{j \in S_i} f_{ij}(\zeta_i, \zeta_j, d_i), \quad e_i = h_i(\zeta_i), \quad i \in {\mathcal I}_N,
\end{eqnarray}
where $e_i$ is the spacing error of the $i^{th}$ vehicle. \textcolor{black}{Please note that the time argument $(t)$ has been dropped for brevity, unless required}.
The equilibrium solution for the above set of coupled evolution equations is  $\zeta_i =0, \; i \in {\mathcal I}_N$. A generalized definition of string stability due to Ploeg et al \cite{PVN2014}, Besselink and Knorn\cite{BK2018} is used here:

\vspace*{0.05in}

\noindent{\textbf{ Definition (Scalable Weak Input-State Stability)}:} The nonlinear system is said to be scalably input-output stable if there exist functions $\beta \in {\mathcal K} {\mathcal L}$ and $\sigma \in {\mathcal K}$ and a number $N_{min}$ such that for any $N \ge N_{min}$ and for any bounded disturbances $d_i(t), \; i \in {\mathcal I}_N$, 
 $$\max_{i \in {\mathcal I}_N}\|\zeta_i(t)\| \le \beta(\sum_{i \in {\mathcal I}_N}\|\zeta_i(0)\|, t) + \sigma(\max_{i \in {\mathcal I}_N} \|d_i(t)\|_{\infty}).$$
\color{black}
The right hand side in the above equation is larger and generalized version of the bound presented in \cite{BK2018}.

We will use the following definition for string stability \cite{strStability_Def_Swaroop2002}:
Suppose a string of vehicles with one-predecessor lookup scheme can be modeled as a set of differential equations of the form:
\begin{align*}
	\dot{x_i}=Ax_i+B\epsilon_{i-1}\\
	\epsilon_{i}=Cx_i+D\epsilon_{i-1},
\end{align*}
where $i=1,2,,\ldots$, and  $x_i(t)\in\mathbb{R}^n \text{ forall } i \text{ and } t$.
Then, a string is considered $L_\infty$ stable if given any $\gamma>0$, there exists a $\delta>0$ such that:
\begin{equation}
	\sup_k \|x_k(0)\|_\infty<\delta \implies\sup_k \|x_k(t)\|_\infty<\gamma.
\end{equation}
The frequency domain condition can then be obtained from this definition \cite{strStability_Def_Swaroop2002} as:
\begin{equation}
	\|H_1(j\omega)\|_\infty \le 1
\end{equation}
\color{black}
\textcolor{black}{With feedback linearization, equation (\ref{eqn:ZetaEOM}) reduces to the equations in Theorem 1}
where $w_0(t)$ denotes the acceleration input of the lead vehicle,
$A_0$ is Hurwitz matrix, $B, C, D$ are respectively constant matrices.


For the platoon of \textcolor{black}{homogeneous} vehicles with one-predecessor lookup scheme, one obtains the following error evolution equations using a Laplace transformation \cite{swaroop1994phd, V2VBenefits}:
$$Y_i(s) = H(s) Y_{i-1}(s), $$
where $H(s)$ is a rational, proper, stable transfer function. The requirement of string stability has thus far \cite{Shahab90, ioannou1993autonomous, swaroop1994phd} been used as $\|H(jw)\|_{\infty} \le 1$.\\

From \cite{desoerfeedback}, it is known that the input-output relationship for a rational, proper transfer function is:
$$\|y_i\|_2 \le \|H(jw)\|_{\infty}\|y_{i-1}\|_2, $$
where the input and output are measured by their ${\mathcal L}_2$ norms (power in the error signals). Practical consideration for this application requires us to consider $\|y_i\|_{\infty}$ (the maximum value of the output) as it has direct bearing on safety. \textcolor{black}{Accordingly}, the corresponding input-output relationship from \cite{desoerfeedback} is
$$\|y_i\|_{\infty} \le \|h(t)\|_1 \|y_{i-1}\|_{\infty}, $$
where $h(t)$ is the unit impulse response of the transfer function $H(s)$. It is known from \cite{desoerfeedback} that $H(0)\le \|H(jw)\|_{\infty} \le \|h(t)\|_1$ and that $H(0) = \int_0^{\infty} h(t) = \|h(t)\|_1$, when $h(t) \ge 0$ for all $t \ge 0$. Typical information flow structures such as the one for ACC and CACC are such that $H(0)=1$, thereby putting a lower bound on $\|h(t)\|_1 =1$. However, for ascertaining string stability, one must attain this lower bound; an obstacle to attaining the lower bound is to find controller gains that render the unit impulse response of $H(s)$ non-negative. This is a variant of the open problem of transient control and there are currently no systematic procedures for determining the set of gains for this case.    

In the first theorem, we exploit bounded structure of leader's acceleration and finite duration of lead vehicle maneuvers to prove that it suffices to consider $\|H(jw)\|_{\infty} \le 1$ to show the {\it uniform} boundedness of spacing errors. \color{black} The first theorem is replicated from \cite{vegamoor2020mobility} for completeness and clarity. The new contribution in this paper is that we have generalized this to multiple vehicle lookup schemes as shown in Theorem \ref{thm:outputBound2Veh}.
\color{black}
\begin{theorem} \label{thm:outputBound}
Suppose:
\begin{itemize}
    \item The  error propagation equations are given by 
    \begin{align}
    \dot \zeta_1(t) &= A_0 \zeta_1(t) + D w_0(t), \label{eqn:Thm1_a}\\
    \dot \zeta_i(t) &= A_0 \zeta_i(t) + B y_{i-1}(t), \quad \forall i \ge 2 \label{eqn:Thm1_b}\\
    y_i(t) &= C \zeta_i(t), \quad \forall i \ge 1 \label{eqn:Thm1_c},
    \end{align}
    and $A_0$ is a Hurwitz matrix; 
    \item the lead vehicle executes a bounded acceleration maneuver in finite time, i.e., $w_0(t) \in {\cal L}_2 \cap {\cal L}_{\infty}$;
    \item $\|C(jw I - A_0)^{-1}B\|_{\infty} \le 1$ and
    \item For some $\alpha^*>0$, $\sum_{i=1}^N \|\zeta_i(0)\| \le \alpha^*$ for every $N$.\textcolor{black}{That is, initially all spacing errors are absolutely summable}.
\end{itemize}
Then, there exists a $M_1, M_2 >0$, independent of $N$, such that for all $i \ge 1$:
$$\|y_{i}(t)\|_{\infty} \le M_1 \alpha^* +M_2\|w_0(t)\|_2. $$
\end{theorem}
\begin{proof}
Since $A_0$ is Hurwitz, we can obtain the following using equations Linear System Theory \cite{desoerfeedback} for some constants $\beta_2, \beta_{\infty}, \gamma_2, \gamma_{\infty}$: 
\begin{align*}
\zeta_1(t) &= e^{A_0t}\zeta_1(0) + \int_0^t e^{A_0(t-\tau)}D w_0(\tau) d \tau, \\
\zeta_i(t) &= e^{A_0t} \zeta_i(0) + \int_0^t e^{A_0(t-\tau)}B y_{i-1}(\tau) d \tau, \;  i \ge 2, \\
\Rightarrow \|y_1(t)\|_2 &\le \beta_2 \|\zeta_1(0)\| + \gamma_2\|w_0(t)\|_2, \\
\|y_1(t)\|_{\infty} &\le \beta_{\infty}\|\zeta_1(0)\| + \gamma_{\infty}\|w_0(t)\|_{\infty}, \\
\|y_i(t)\|_2 &\le \beta_2 \|\zeta_i(0)\| +  \|y_{i-1}(t)\|_2, \; \; i \ge 2.
\end{align*}
Note that the last inequality results from $ \|C(jwI-A_0)^{-1}B\|_{\infty} \le 1$. These inequalities can be combined as:
\begin{align*}
\|y_{i}(t)\|_2 &\le \beta_2 (\sum_{j=2}^i \|\zeta_j(0)\|) + \|y_1(t)\|_2, \\
&\le \beta_2 (\sum_{i\in {\cal I}_N} \|\zeta_i(0)\|) + \gamma_2 \|w_0(t)\|_2, \\
&\le \beta_2 \alpha^* + \gamma_2 \|w_0(t)\|_2.
\end{align*}

\textcolor{black}{Corless et al \cite{corless-zhu-skelton} have provided bounds on the $L_\infty$ norm of the output, given an $L_2$ input for an asymptotically stable system. Using their approach, it follows that if we can optimize a scalar $J$:}
\begin{align*}
	J := \min \{g\}, \quad \textcolor{black}{\text{subject to:}} \\
	C^TPC-gI\prec 0, \text{\textcolor{black}{ $g\in \mathbb{R^+}$}}\\
	P \succ 0, \quad A_0 P + PA_0^T+BB^T = 0,
\end{align*}
then for some $\eta >0$ and for all $i\ge 1$,
\begin{eqnarray*}
\|y_{i}(t)\|_{\infty} &\le&  \eta \|\zeta_i(0)\| + \sqrt{J} \|y_{i-1}(t)\|_2, \\
&\le& (\sqrt{J} \beta_2+\eta) \alpha^* + \sqrt{J} \gamma_2\|w_0(t)\|_2.
\end{eqnarray*}
This completes the proof.
\end{proof}
\noindent{\textbf{Remark}:} If the platoon has a finite number of vehicles then the last condition on the sum of absolute initial errors can be trivially satisfied. Since longitudinal maneuvers of a vehicle change the speed of a vehicle from a constant to another constant in finite time, the lead vehicle's exogenous input $w_0(t)$ (e.g., its acceleration or jerk) can be assumed without any loss of generality to be square integrable. 
\textcolor{black}{Now, for a two vehicle lookup scheme, the requirement placed on the transfer functions is}:
\begin{equation} \label{eqn:SumOfTFs}
\|H_1(j\omega)\|_\infty+\|H_2(j\omega)\|_\infty\le1,
\end{equation}
where:
\begin{align*}
\|H_1(s)\|_\infty=\frac{Y_i(s)}{Y_{i-1}(s)}\\
\|H_2(s)\|_\infty=\frac{Y_i(s)}{Y_{i-2}(s)}\\
\end{align*}
Consequently, we obtain an extension of Theorem \ref{thm:outputBound} for a CACC+ policy:
\begin{theorem} \label{thm:outputBound2Veh}
	Suppose the  error propagation equations are given by 
	\begin{align}
	\dot \zeta_1(t) &= A_0 \zeta_1(t) + D w_0(t), \\
	\dot \zeta_2(t) &= A_0 \zeta_2(t) + B_1 y_1(t), \\
	\forall i \ge 3, \quad \dot \zeta_i(t) &= A_0 \zeta_i(t) + B_1  y_{i-1}(t)+B_2y_{i-2}(t),  \\
	\forall i \ge 1, \quad y_i(t) &= C \zeta_i(t),  
	\end{align}
	where $A_0$ is a Hurwitz matrix; 
	furthermore, suppose that 
	\begin{itemize}
		\item the lead vehicle executes a bounded acceleration maneuver in finite time, i.e., $w_0(t) \in {\cal L}_2 \cap {\cal L}_{\infty}$;
		\item $\|C(jw I - A_0)^{-1}B_1\|_{\infty} +\|C(jw I - A_0)^{-1}B_2\|_{\infty} \le 1$ and
		\item For some $\alpha^*>0$, $\sum_{i=1}^N \|\zeta_i(0)\| \le \alpha^*$ for every $N$. 
	\end{itemize}
Then, there exists a $M_1, M_2 >0$, independent of $N$, such that for all $i \ge 1$:
	$$\|y_{i}(t)\|_{\infty} \le M_1  +M_2\|w_0(t)\|_2. $$
\end{theorem}
\begin{proof}
	For some constants $\beta_2, \beta_{\infty}, \gamma_2, \gamma_{\infty}$, we obtain: 
	\begin{align*}
	\zeta_1(t) &= e^{A_0t}\zeta_1(0) + \int_0^t e^{A_0(t-\tau)}D w_0(\tau) d \tau, \\
	\zeta_2(t) &= e^{A_0t}\zeta_2(0) +  \int_0^t e^{A_0(t-\tau)}B_1 y_{1}(\tau) d \tau , \\
	\zeta_i(t) &= e^{A_0t} \zeta_i(0) + \int_0^t e^{A_0(t-\tau)}B_1 y_{i-1}(\tau) d \tau \\ & \qquad +\int_0^t e^{A_0(t-\tau)}B_2 y_{i-2}(\tau) d \tau, \;  i \ge 2, \\
	\Rightarrow \|y_1(t)\|_2 &\le \beta_2 \|\zeta_1(0)\| + \gamma_2\|w_0(t)\|_2, \\
	\|y_1(t)\|_{\infty} &\le \beta_{\infty}\|\zeta_1(0)\| + \gamma_{\infty}\|w_0(t)\|_{\infty}, \\
	\|y_2(t)\|_2 &\le \beta_2 \|\zeta_2(0)\| +  \|y_1(t)\|_2.
	\end{align*}
	Using equation (\ref{eqn:SumOfTFs}):
	\begin{align*}
	&\|y_i(t)\|_2 \\
	&\le \beta_2 \|\zeta_i(0)\| + \|C(jwI-A_0)^{-1}B_1\|_{\infty} \|y_{i-1}(t)\|_{2} \\ 
	& \quad+\|C(jwI-A_0)^{-1}B_2\|_{\infty} \|y_{i-2}(t)\|_{2} \\
	&\le \beta_2 \|\zeta_i(0)\|  \\
	&\quad+ \|C(jwI-A_0)^{-1}B_1\|_{\infty}\max \{\|y_{i-1}(t)\|_2, \|y_{i-2}(t)\|_{2} \} \\
	& \qquad+ 
	\|C(jwI-A_0)^{-1}B_2\|_{\infty}\max \{\|y_{i-1}(t)\|_2, \|y_{i-2}(t)\|_{2} \}\\
	&\le \beta_2 \|\zeta_i(0)\| + \max\{\|y_{i-1}(t)\|_2, \|y_{i-2}(t)\|_2 \} \; \; \forall i \ge 3.
	\end{align*}
	\textcolor{black}{After simplification},
	\begin{align*}
	\|y_i(t)\|_2 &\le \beta_2 \sum_{j=3}^i  \|\zeta_i(0)\| + \max \{\|y_1(t)\|_2, \|y_2(t)\|_2\} \\
	&\le \beta_2 \sum_{j=1}^i \|\zeta_i(0)\| + \gamma_2 \|w_0(t)\|_{2}
	\end{align*}
	From \cite{corless-zhu-skelton}, it follows that if 
	\begin{align*}
	J &:= \min \{g\}, \quad \textcolor{black}{\text{subject to:}} \\
	&C^TPC-gI\prec 0,\\
	P \succ 0, \quad &A_0P + PA_0^T+B_1B_1^T+B_2B_2^T = 0,
	\end{align*}
	then for some $\eta >0$ and for all $i\ge 1$,
	\begin{align*}
	\|y_{i}(t)\|_{\infty} &\le  \sqrt{J}\left|\left|\begin{matrix}
	y_{i-1}\\y_{i-2}
	\end{matrix}\right|\right|_2\\
	&\le \sqrt{J}(\|y_{i-1}(t)\|_2 + \|y_{i-2}(t)\|_2) \\
	&\le 2 \sqrt{J} (\beta_2 \sum_{j=1}^i \|\zeta_i(0)\|+ \gamma_2 \|w_0(t)\|_2) \\
	& \le M_1  +M_2\|w_0(t)\|_2,
	\end{align*}
	by setting $M_1 = 2 \sqrt{J} \beta_2 \alpha^*$ and $M_2 = 2 \sqrt{J} \gamma_2$. 
	This completes the proof. One requires $w_0(t) \in {\cal L}_{\infty}$  to guarantee that $\|y_i(t)\|_{\infty}$ is bounded. 
\end{proof}

\noindent{\textbf{Remark}:} Theorem \ref{thm:outputBound2Veh} can also be applied to a platoon with n-vehicle lookup scheme by modifying the string stability criterion to:
\begin{align*}
\sum_{k=1}^{n}\|H_k(j\omega)\|_\infty&\le 1 \\
\text{i.e.,  }\sum_{k=1}^{n}\|C(jw I - A_0)^{-1}B_k\|_{\infty}  &\le 1
\end{align*}
%
%

\section{Lossy One-Predecessor Lookup Scheme} \label{sec:lossyCACC}
\subsection{Gilbert Noise Channel}\label{sec:GilbertIntro}
\begin{figure}[htbp]
	\centering
	\includegraphics[width=0.45\textwidth]{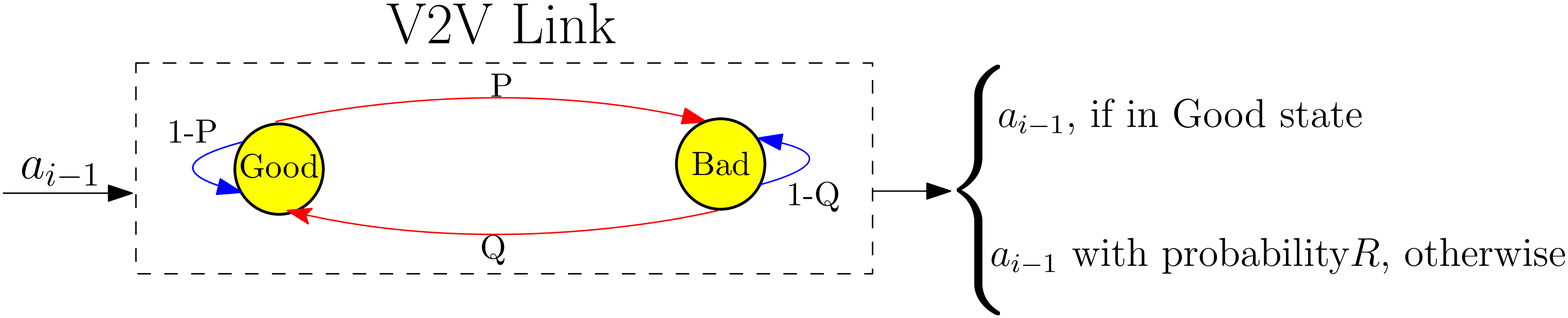}
	\caption{Transmission diagram for wireless link between the $i^{th}$ and $(i-1)^{th}$ vehicle.}
	\label{fig:GilbertFig}
\end{figure}

A burst-noise channel model for bit errors was introduced in 1960 by E. N. Gilbert \cite{Gilbert1960}. Since then, it has been widely used (along with some of the model's extensions \cite{Gilbert_Eliott, ExtendedGilbert}) to simulate bursts of noise and packet drops in wireless channels. In its simplest form, the Gilbert model can be described as a two-state system as shown in Fig. \ref{fig:GilbertFig}. The input to the wireless channel in our case is the acceleration of the preceding vehicle. In the `Good' state, there are no packets dropped and the input information is transmitted successfully. In the `Bad' state, only $R\%$ of the packets are transmitted successfully. Further, the transition probabilities from `Good'$\rightarrow$`Bad' and `Bad'$\rightarrow$`Good' are $P$ and $Q$ respectively. $P$ and $Q$ are typically small so the states tend to persist for a few transmission cycles, imparting burst behavior to the channel.
Consequently, we can represent the net output of the wireless channel as $\hat{w}a_{i-1}$, where $\hat{w} \in \{0,1\}$ with $\hat{w}=1$ when the packet is transmitted successfully. Since errors only occur in the dropped state, we can calculate the average packet reception rate which we denote as $\gamma$.

\begin{align}\label{eqn:GilbertGamma}
	\mathbb{E}[\hat{w}]=1-\frac{P(1-R)}{P+Q} = :\gamma,
\end{align}
\color{black}
It should be noted that in practice, average packet reception rate, $\gamma$ can be locally measured by the receiver if transmission rate is fixed (as can be expected from real-time systems) and known before hand (either by design or as part of an eventual V2V standard). The choice of interval used for updating $\gamma$ can be left to the designer based on the transmission bandwidth available.
\color{black}

If the communication link were ideal, then the control law for a one-predecessor lookup scheme (CACC) can be written as a deterministic equation:
\begin{align}
        u_i=K_{a}a_{i-1}-K_{v}(v_i-v_{i-1})-K_{p}(x_i-x_{i-1}+h_wv_i).
\end{align}
Here, $x_i$, $v_i$, $a_i$ are the position, velocity and acceleration states of the $i^{th}$ vehicle. The constants $K_a, K_v$ and $K_p$ are controller gains, $h_w$ is the chosen time headway, and $u_i$ is the control input for the $i^{th}$ vehicle. On the other hand, with a lossy V2V link, the equations of motion for each of the $i^{th}$ follower vehicles will be stochastic:
\begin{align} \label{eqn:StochasticCACC}
	\tau \dot{a}_i+a_i&=u_i=\hat{w}_{i,i-1}K_{a}a_{i-1}-K_{v}(v_i-v_{i-1}) \nonumber \\
	& \qquad \qquad-K_{p}(x_i-x_{i-1}+h_wv_i).
\end{align}
Here, $\hat{w}_{i,i-1} \in \{1,0\}$ is a binary random variable which takes the value $1$ when the acceleration packet is successfully transmitted from the $(i-1)^{th}$ vehicle to the $i^{th}$ vehicle and takes value $0$ otherwise.

Since it is well known that losses in the V2V link can affect string stability\cite{LeiITSconf2011,LossyCommsVargas}, it is desirable to obtain a relation between the minimum string stable time headway and the loss characteristics of the channel. Such a relation was derived in our previous work \cite{vegamoor2019} but required that the packet losses be i.i.d (independent, identically distributed) in nature. In this paper, the i.i.d assumption \textcolor{black}{between packets within each V2V link} is relaxed and we derive a sufficient condition for the time headway which is applicable for burst-type losses as well. The i.i.d case is in fact a special case of the Gilbert model with only a `Bad' state where successful packet reception occurs with probability $\gamma=R$.

\subsection{Convergence of State Vector for Lossy CACC}\label{subSec:GilbertCACC}
First, we will show using induction that when a large number of realizations of the stochastic system (\ref{eqn:StochasticCACC}) are taken, the average approaches a known deterministic equivalent. This deterministic system will then be used to obtain the minimum string stable time headway. This induction procedure for lossy CACC has been duplicated from our recent conference paper \cite{vegamoor2020mobility} for completeness as we will refer to it again for the CACC+ case in section \ref{sec:lossyCACC+}.

We will operate under the assumption that the communication link between any pair of vehicles is independent from any other pair. \textcolor{black}{This is largely true in non-congested scenarios. Admittedly, this simplification was necessary to maintain tractability of the problem. Moreover, similar assumptions of independence have been used successfully by other researchers \cite{independentV2VEg1,JointCommDelay}.}

Suppose there are $k$ vehicles in the platoon operating with the stochastic control law in (\ref{eqn:StochasticCACC}). We can consider $\gamma$ for the platoon to be the smallest among those measured over the individual V2V links. Let  $u_L$ be the input imparted to the lead vehicle. The stochastic system of the entire platoon can be written in the state space form as:
\begin{align}\label{eqn:StochasticStateSpace}
    \dot{\hat{X}}=\hat{A}(\hat{w}(t))\hat{X}+BU,
\end{align}
where $\hat{X}$=$(x_0,v_0,a_0,x_1,v_1,a_1,\cdots x_k,v_k,a_k)$ and $U =u_L$, the input to the lead vehicle. Note that only the system matrix $\hat{A}(\hat{w}(t))$ has random elements. For sake of clarity, we have provided an example of $\hat{A}$ and $B$ using equation (\ref{eqn:StochasticCACC}) for a three (1 lead, 2 following) vehicle CACC platoon with imperfect (lossy) V2V communication. The two random entries in the $9\times 9$ system matrix are highlighted.

 \renewcommand\arraystretch{1}
 \setlength{\arraycolsep}{2.5pt}
 \begin{align}\label{eqn:SS_CACC}
     &\hat{A}_{L}=\begin{bmatrix}\hspace {0.2em} 0&1&0& & & & & &\\
     0&0&1 &&&&&&\\
     0&0&\frac{-1}{\tau}&&&&&&&\\
     &&&0&1&0&&&&\\
     &&&0&0&1&&&&\\
     \frac{K_{p}}{\tau}&\frac{K_{v}}{\tau}&\textcolor{orange}{\bm{\frac{\hat{w}_{1,0}K_{a}}{\tau}}}&\frac{-K_{p}}{\tau}& p_1&\frac{-1}{\tau}&&&\\
     &&&&&&0&1&0&\\
     &&&&&&0&0&1&\\
     0&0&0&\frac{K_{p}}{\tau}&\frac{K_{v}}{\tau}&\textcolor{orange}{\bm{\frac{\hat{w}_{2,1}K_{a}}{\tau}}}&\frac{-K_p}{\tau}{}&p_1&\frac{-1}{\tau}
     \end{bmatrix},\nonumber \\
     &B_{L}=\begin{bmatrix} 0&0&1&0&0&0&0&0&0 \end{bmatrix}^T,
 \end{align}
 \renewcommand\arraystretch{1}
 where
 \begin{align}
     p_1=-\frac{K_{v}+K_{p}h_w}{\tau}. \nonumber
 \end{align}

Let $\Delta t$ be the controller time step so that the total (finite) run time is $t_m =m\Delta t$, $m\in \mathbb{N}$. Let us consider the evolution of the stochastic state vector over the first time interval $[0,t_1)$:
\begin{align}
    \hat{X}(t_1)= \hat{\Phi}(t_1,0)\hat{X}(0)+\int_{0}^{t_1}\hat{\Phi}(t_1,\zeta)BU(\zeta)d\zeta \label{eqn:StochasticSTM},
\end{align} 
where ${\hat{\Phi}(t_1,0)}$ is the stochastic state transition matrix, dependent on the values of $\hat{w}_{i,j}$ at $t=0$.  In implementation, the control input $U$ would only be updated at the beginning of each time step so it can be considered constant over the duration of each interval.
 
Since we have defined $\mathbb{E}[\hat{w}_{i,j}]=\gamma$, let us consider replacing the random elements in the system matrix of equation (\ref{eqn:StochasticStateSpace}) with their expected values. Then we get some deterministic system:

\begin{align}
        \dot{\bar{X}}=\bar{A}\bar{X}+BU
\end{align}

Our goal now is to show that $\mathbb{E}[\dot{\hat{X}}(t)]=\dot{\bar{X}}(t)$, for all $t \in [0,t_m]$.
For the deterministic system, the state evolution for the first interval $[0,t_1)$ is:
\begin{align}\label{eqn:DeterministicSTM}
     \bar{X}(t_1)= \bar{\Phi}(t_1,0)\bar{X}(0)+\int_{0}^{t_1}\bar{\Phi}(t_1,\zeta)d\zeta BU(0).
\end{align}

Now consider $\hat{\Phi}(t_1,0)$ and $\bar{\Phi}(t_1,0)$. Since $\hat{A}(\hat{w}(t))$ only changes at each controller time step, it is constant in the interval $[0,t_1)$ and \textcolor{black}{retains the value from the beginning of the interval} $\hat{A}(\hat{w}(0))=:\hat{A}_1$. So, we can write
\begin{align}
        \hat{\Phi}(t_1,0)&=e^{\int_0^{t_1} \hat{A}(\hat{w}(\xi))d\xi}= e^{\hat{A}_1t_1} \label{eqn:BasePhiHat}\\
        \bar{\Phi}(t_1,0)&=e^{\int_0^{t_1} \bar{A}d\xi}= e^{\bar{A}t_1}
\end{align}

Now we use the power series expansion for the exponential matrices:
\begin{align}
    e^{\hat{A}_1t_1}&=I+\hat{A}_1t_1+\frac{(\hat{A}_1t_1)^2}{2!}+\frac{(\hat{A}_1t_1)^3}{3!} +\cdots \label{eqn:expAhat}\\ 
    e^{\bar{A}t_1}&=I+\bar{A}t_1+\frac{(\bar{A}t_1)^2}{2!}+\frac{(\bar{A}t_1)^3}{3!} +\cdots \label{eqn:expAbar}
\end{align}

While generally not true for random matrices \cite{matrixConvexity}, the following is true for CACC system matrices with one vehicle lookup case given in equation (\ref{eqn:SS_CACC}):
\color{black}
\begin{theorem}
\begin{equation}\label{eqn:MatrixExpectation2}
\mathbb{E}[\hat{A}^n_1]= \bar{A}^n, \quad \forall n \in \mathbb{N}
\end{equation}
\end{theorem}
Proof of this theorem is given in the Appendix.
Proof for this theorem holds due to its specific structure since the diagonal elements of the system matrix are purely deterministic and the powers of $\hat{A}$ only contain elements that are multi-linear in $\hat{w}_{i,j}$. \color{black}
This allows us to exploit the fact that the expectation of a product of independent random variables is the product of their expectations. We have noticed that this convenient multi-linear property of the powers of system matrices is afforded only for one vehicle lookup schemes (CACC) but not for platoons that utilize communicated information from two or more preceding vehicles (CACC+ systems), as discussed in Section \ref{sec:lossyCACC+}.

Thus, over a large number of realizations,
    \begin{align}\label{eqn:BasePhiExp}
        \mathbb{E}[\hat{\Phi}(t_1,0)]=\bar{\Phi}(t_1,0). 
    \end{align}
Since the initial conditions can be assumed to be the same in equations (\ref{eqn:StochasticSTM}) and (\ref{eqn:DeterministicSTM}), i.e., $\hat{X}(0)=\bar{X}(0)$, we get:
\begin{align}
    \mathbb{E}[\hat{X}(t_1)]=\bar{X}(t_1),
\end{align} \label{eqn:InductionBase}
for the first interval $[0,t_1)$. Let this form the base case with the induction hypothesis for interval $[t_{k-1},t_k)$ as:
\begin{align}
    \mathbb{E}[\hat{X}(t_k)]=\bar{X}(t_k) \label{eqn:InductionHypth}
\end{align}
Now consider the next interval $[t_k,t_{k+1})$
\begin{align*}
    \hat{X}(t_{k+1})= \hat{\Phi}(t_{k+1},t_k)\hat{X}(t_k)+\int_{t_k}^{t_{k+1}}\hat{\Phi}(t_{k+1},\zeta)d\zeta BU(t_k) \\
    \bar{X}(t_{k+1})= \bar{\Phi}(t_{k+1},t_k)\bar{X}(t_k)+\int_{t_k}^{t_{k+1}}\bar{\Phi}(t_{k+1},\zeta)d\zeta BU(t_k)
\end{align*}
Using a similar reasoning as in equations \eqref{eqn:BasePhiHat} - \eqref{eqn:BasePhiExp}, we can show that $\mathbb{E}[\hat{\Phi}(t_{k+1},t_k)]=\bar{\Phi}(t_{k+1},t_k)$.

Again, note that the term $\hat{\Phi}(t_{k+1},t_k)\hat{X}(t_k)$ only contains products of independent random variables. From the induction hypothesis in equation (\ref{eqn:InductionHypth}), we can claim $\mathbb{E}[\hat{\Phi}(t_{k+1},t_k)\hat{X}(t_k)]=\bar{\Phi}(t_{k+1},t_k)\bar{X}(t_k)$. This yields:
\begin{align}
    \mathbb{E}[\hat{X}(t_{k+1})]=\bar{X}(t_{k+1}).
\end{align}
From the principle of mathematical induction, $\mathbb{E}[\hat{X(t)}]=\bar{X}(t)$ for all finite $t \in [0,t_m]$. This allows us to replace equation (\ref{eqn:StochasticCACC}) with its deterministic equivalent.
\begin{align}
    \tau \dot{a}_i+a_i&=\gamma K_{a}a_{i-1}-K_{v}(v_i-v_{i-1}) \nonumber \\ & \qquad-K_{p}(x_i-x_{i-1}+d+h_w v_i) 
\end{align}
Following the procedure in \cite{V2VBenefits} for this governing equation, we obtain the bound on the minimum employable time headway.
\begin{align} \label{eqn:PL_CACC_Limit}
h_w\ge h_{min}=\frac{2\tau}{1+\gamma K_a} 
\end{align}

\section{Approximate Convergence of State Vector for Two-Predecessor Lookup}\label{sec:lossyCACC+}
Now let us consider a two vehicle lookup scheme with packet losses. The equation of motion for each vehicle in the platoon is given by:
\begin{align} 
\tau \dot{a}_0+a_0&=u_L \label{eqn:2VehEOMStart}\\
\tau \dot{a}_1+a_1&=\hat{w}_{1,0}K_{a}a_0-K_{v}(v_1-v_0) \nonumber\\ 
& \qquad -K_{p}(x_1-x_0+d+h_wv_1)\\
\tau \dot{a}_i+a_i&=\hat{w}_{i,i-1}K_{a}a_{i-1}-K_{v}(v_i-v_{i-1}) \nonumber\\
& \qquad -K_{p}(x_i-x_{i-1}+d+h_wv_i) \nonumber\\
&\qquad +\hat{w}_{i,i-2}\{K_{a}a_{i-2}-K_{v}(v_i-v_{i-2}) \nonumber\\
&\qquad \qquad -K_{p}(x_i-x_{i-2}+2d+2h_wv_i)\},\; \forall i\ge 2. \label{eqn:2VehEOMEnd}
\end{align}

The above system is stochastic, and we would like to obtain a deterministic equivalent of the system in order to derive a sufficient condition for a string stable time headway that can be deployed over lossy communication channels. The stochastic system can be expressed in a similar state space form  as in equation (\ref{eqn:StochasticStateSpace}). For clarity, $\hat{A}(\hat{w}(t))$ is provided for a (2+1) vehicle platoon, with the last vehicle using acceleration information from the second and full state information from leading vehicle. It should be stressed that unlike the CACC case, the position and velocity information of the $(i-2)^{th}$ vehicle cannot be measured using onboard sensors on the ego ($i^{th}$) vehicle, so they are transmitted wirelessly along with the acceleration information. Again, the random entries in the matrix are highlighted.
\begin{align}
 \setlength{\arraycolsep}{0.5pt}
A_2&= \nonumber\\
& \begin{bmatrix} 0&1&0& & & & & &\\
0&0&1 &&&&&&\\
0&0&\frac{-1}{\tau}&&&&&&&\\
&&&0&1&0&&&&\\
&&&0&0&1&&&&\\
\frac{K_{p}}{\tau}&\frac{K_{v}}{\tau}&\textcolor{orange}{\frac{\hat{w}_{1,0}K_{a}}{\tau}}&\frac{-K_{p}}{\tau}& P_1&\frac{-1}{\tau}&&&\\
&&&&&&0&1&0&\\
&&&&&&0&0&1&\\
\textcolor{orange}{\frac{\hat{w}_{2,0}K_{p}}{\tau}}&\textcolor{orange}{\frac{\hat{w}_{2,0}K_{v}}{\tau}}&\textcolor{orange}{\frac{\hat{w}_{2,0}K_{a}}{\tau}}&\frac{K_{p}}{\tau}&\frac{K_{v}}{\tau}&\textcolor{orange}{\frac{\hat{w}_{2,1}K_{a}}{\tau}}&\textcolor{orange}{P_2}&\textcolor{orange}{P_3}&\frac{-1}{\tau}
\end{bmatrix} \nonumber\\
B_{L}&=\begin{bmatrix} 0&0&1&0&0&0&0&0&0 \end{bmatrix}^T,
\end{align}
where
\begin{align}
P_1&=-\frac{K_{v}+K_{p}h_w}{\tau} \nonumber \\
P_2&=-\frac{K_{p}+\hat{w}_{2,0}K_{p}}{\tau} \nonumber \\
P_3&=-\frac{K_{v}+K_{p}h_w+\hat{w}_{2,0}(K_{v}+2K_{p}h_w)}{\tau}
\end{align}	 
Suppose we attempt to use a similar approach to that presented in section \ref{subSec:GilbertCACC}, we see that the powers of $\hat{A}_2$ matrix are no longer multi-linear in the random elements. Consequently,
\begin{equation*}\label{eqn:MatrixExpectationNotEqual}
\mathbb{E}[\hat{A_2}^n]\ne \bar{A_2}^n, \qquad \forall n \ge 3, 
\end{equation*}
 which renders the earlier approach futile.

The task of obtaining an exact deterministic equivalent of \crefrange{eqn:2VehEOMStart}{eqn:2VehEOMEnd} in essence, can be represented as follows: 

Given a random matrix $S$ whose elements are not necessarily independent of each other, find a deterministic matrix $D$ such that:
\begin{align} \label{eqn:exponentialRandomMat}
\mathbb{E}[e^S]=e^D
\end{align}

To the best of our knowledge, finding an exact expression for $D$ appears to be tedious for non-trivial cases. While a wealth of results are available in random matrix theory, they either rely on diagnonalizability of the matrix or independence of its elements \cite{RandomProd1,RandomProd2}. S. Geman and R. Khasminskii \cite{StuartG1,Khasminskii2008} provide some results on convergence of stochastic differential equations, but they appear to require infinitesimally small time steps, which is not practical for implementation on real vehicles. A brute force computational method can be pursued where the matrix exponential of a large number of realizations of the $S$ matrix are taken and averaged to get $e^D$. Then its matrix logarithm needs to be calculated numerically to obtain $D$. We observed a significant loss of precision due to the multiple floating point operations involved in taking matrix exponentials. This causes difficulty in finding a real valued matrix logarithm.

\textcolor{black}{While finding an exact deterministic expression is difficult, we propose the following approximation for the system, by replacing all random variables with their expectations}. 
\begin{align} 
\tau \dot{a}_0+a_0&=u_L \label{eqn:2VehEOMGammaStart}\\
\tau \dot{a}_1+a_1&=\gamma K_{a}a_0-K_{v}(v_1-v_0) \nonumber\\ 
& \qquad -K_{p}(x_1-x_0+d+h_wv_1)\\
\tau \dot{a}_i+a_i&=\gamma K_{a}a_{i-1}-K_{v}(v_i-v_{i-1}) \nonumber\\
& \qquad -K_{p}(x_i-x_{i-1}+d+h_wv_i) \nonumber\\
&\qquad +\gamma \{K_{a}a_{i-2}-K_{v}(v_i-v_{i-2}) \nonumber\\
&\qquad \qquad -K_{p}(x_i-x_{i-2}+2d+2h_wv_i)\},\; \forall i\ge 2 \label{eqn:2VehEOMGammaEnd}
\end{align}

\begin{figure}[htbp]
	\centering
	\includegraphics[width=9.5cm]{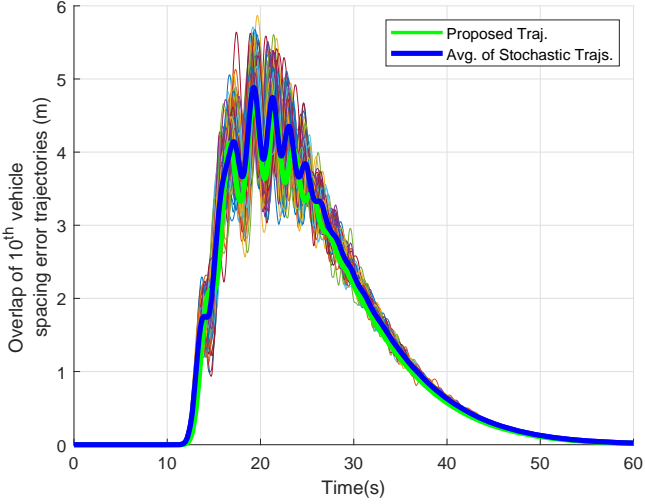}
	\caption{\textcolor{black}{Overlaid stochastic spacing error trajectories and the proposed approximation}}
	\label{fig:Envelop_Pos}
\end{figure}

To understand the behavior of this approximated deterministic system compared to the stochastic system, we simulated 100 realizations of a 10 vehicle stochastic platoon during an emergency braking scenario. \color{black} Time step used in all the simulations was $0.01$ \textcolor{black}{s}. Transition probabilities for the Gilbert channel from Fig \ref{fig:GilbertFig} were picked as $P=0.2, Q=0.1, R=0.2$. All 100 stochastic spacing error trajectories of the $10^{th}$ vehicle are shown in Fig. \ref{fig:Envelop_Pos}, along with their average. The corresponding spacing error trajectory from the proposed system is also shown in the same figure. While we know that \crefrange{eqn:2VehEOMGammaStart}{eqn:2VehEOMGammaEnd} are not the deterministic equivalent of \crefrange{eqn:2VehEOMStart}{eqn:2VehEOMEnd}, we can see that the difference in peaks between the proposed and average trajectories is relatively small (in this case, $0.21$ \textcolor{black}{m}, less than $5\%$ of the peak value). \color{black}So for the purpose of developing an analytical bound for the minimum string stable time headway, we will proceed with \crefrange{eqn:2VehEOMGammaStart}{eqn:2VehEOMGammaEnd}.
\color{black}
We can first define the spacing error for the $i^{th}$ vehicle as:
\begin{equation} \label{eqn:SpacingErrorDef}
	e_i:=x_i-x_{i-1}+d+h_w v_i,
\end{equation}
where $x_i$ is the position of the $i^{th}$ vehicle.
\color{black}
After some algebraic manipulation, we obtain the following equation of motion for each of the $i^{th}$ following vehicle, for $i\ge2$:

\begin{align}\label{eqn:CACC+}
\tau \dddot{e}_i+\ddot{e}_i=&\quad K_p e_{i-1} - K_p e_i - K_v \dot{e}_i -\gamma  K_v\dot{e}_i - K_p h_w \dot{e}_i \nonumber \\& \quad +K_v \dot{e}_{i-1}+ \gamma  K_p e_i+\gamma K_a \ddot{e}_{i-1}+\gamma K_a \ddot{e}_{i-2} \nonumber\\&\quad+\gamma  K_v \dot{e}_{i-2} +\gamma  K_p e_{i-2} +2 \gamma  K_p h_w \dot{e}_i 
\end{align}

This can be written in the Laplace domain as:

\begin{align}
E_i (s) =H_{p1}E_{i-1} (s) + H_{p2}E_{i-2} (s)
\end{align}

where 

\begin{align}
&H_{p1}(s)\nonumber\\&=\frac{\gamma K_a s^2 +K_v s +K_p}{\tau s^3 +s^2 +s[(1+\gamma )K_v+(1+2\gamma )K_ph_w]+(1+\gamma )K_p}
\end{align}

and

\begin{align}
&H_{p2}(s)\nonumber \\&= \frac{\gamma K_a s^2 +\gamma  K_v s+\gamma  K_p}{\tau s^3 +s^2 +s[(1+\gamma ) K_v+(1+2\gamma )K_ph_w]+(1+\gamma )K_p}
\end{align}

We can obtain minimum required time headway $h_{min}$ for the lossy CACC+ platoon by taking the maximum of the two yielded from $H_{p1}(s)$ and  $H_{p2}(s)$, following the method in \cite{V2VBenefits}. Thus, the sufficient condition on time headway for two vehicle lookup is:
\begin{align} \label{eqn:CACC+_hwLimit}
h_w \geq h_{min}=\quad \frac{2\tau (1+\gamma)}{(1+2\gamma )(1+\gamma(1+\gamma )K_a)}
\end{align}
\color{black}

Please note that in equations \crefrange{eqn:2VehEOMGammaStart}{eqn:2VehEOMGammaEnd}, we have assumed that the expectations are all equal to $\gamma$. It is possible that $\hat{w}_{2,0}$ may have a significantly different value from that of $\hat{w}_{1,0}$ or $\hat{w}_{2,1}$ due to poor line-of-sight (since there is a vehicle in between). In that case, we can consider $\mathbb{E}[\hat{w}_{2,0}]=\mu$ and $\mathbb{E}[\hat{w}_{1,0}]=\mathbb{E}[\hat{w}_{2,1}]=\gamma$. The latter equality can be enforced by picking the minimum of the individually measured packet reception rates, which should be close to each other since they both have direct line of sight of the preceding vehicle.
Then, the limit can be obtained from similar algebra:
\begin{align} \label{eqn:CACCPlushwLimitMu}
	h_w \geq h_{min}=\quad \frac{2\tau (1+\gamma)}{(1+2\mu )(1+\gamma(1+\mu )K_a)}
\end{align}

\color{black}

%
\FloatBarrier
\section{Simulations} \label{sec:SimulationMainSec}
The ideal method to corroborate the bound on minimum time headway is to implement the controller on four or more passenger cars and perform real-world experiments, which is logistically demanding. Also, it would be expensive to demonstrate string instability under emergency braking scenarios with real vehicles. Instead, we develop longitudinal model of a 2017 Lincoln MKZ using throttle and brake maps. Once the model is validated using experimental data, we implement six virtual vehicles in Simulink to demonstrate the advantages of the proposed algorithm. As a preliminary check, we first perform simulations using the linear point mass model from equation (\ref{eqn:PointMassModel}).

\subsection{Preliminary Simulations with Point Mass model}\label{sec:LinearModelSims}
We will simulate a homogeneous platoon with actuation lag $\tau=0.4s$ using Simulink. Transition probabilities for the Gilbert channel from Fig. \ref{fig:GilbertFig} were picked as: $P=0.2, Q=0.1, R =0.2$. This yields $\gamma = 0.467$ from equation (\ref{eqn:GilbertGamma}). Here, we will consider a CACC+ platoon of seven (one lead + six follower) vehicles using the control scheme given in \crefrange{eqn:2VehEOMStart}{eqn:2VehEOMEnd}. For simulations with the same linear model with one-vehicle lookup, please refer to \cite{vegamoor2019,vegamoor2020mobility}. For CACC+, the first following vehicle only has one predecessor so it uses the CACC control law from equation (\ref{eqn:StochasticCACC}).

\begin{figure}[htbp] 
	\centering
	\includegraphics[height=10cm]{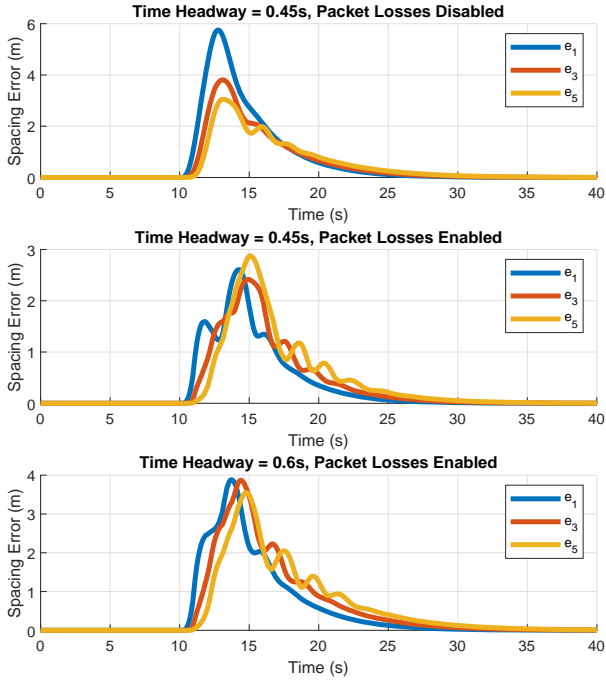}
	\caption{Spacing errors a CACC+ platoon with linear point mass model under different scenarios}
	\label{fig:GilbertLinearCombined}
\end{figure}

The lead vehicle's maneuver is as follows: At the start of the simulation, it is moving with a velocity of $25$ \textcolor{black}{m/s}, then at $t=10$ \textcolor{black}{s}, it decelerates at the rate of $-9$ \textcolor{black}{m/s$^2$} to $16$ \textcolor{black}{m/s}. This velocity is maintained for the remaining duration of the simulation. \textcolor{black}{In all the following simulations, we assume that the platoon is in steady state at the start of each run.} This setup is similar to an emergency braking maneuver. Controller gains were chosen as follows: $(K_a, K_v, K_p) =(0.2,2.5,1)$. Spacing error trajectories for the $1^{st}$, $3^{rd}$ and $5^{th}$ follower vehicles for three different communication scenarios are presented in Fig. \ref{fig:GilbertLinearCombined}. 

First, the platoon is simulated with a time headway of $0.45$ \textcolor{black}{s} but with no packet losses. This scenario is expected to result in a string stable platoon, since the headway exceeds the minimum bound of $0.38$ \textcolor{black}{s} from \cite{V2VBenefits}. In the next scenario, the platoon uses the same time headway but packet losses are enabled using the Gilbert channel described earlier. \textcolor{black}{This can be confirmed visually as well since the $L_\infty$ norms are in the order: $\|e_1\|_\infty \ge \|e_3\|_\infty \ge \|e_5\|_\infty$. That is, the maximum spacing errors diminish across the platoon. Please note that the maxima for each curve is that over the entire simulation interval, as per the $L_\infty$ definition of string stability mentioned in Section \ref{sec:SpacingBound}}. We can see from the second subplot in Fig. \ref{fig:GilbertLinearCombined} that maintaining the same time headway induces string instability, since the last follower's \textcolor{black}{maximum} spacing error is larger than that of the first. Finally, since equation (\ref{eqn:CACC+_hwLimit}) yields a minimum value of $0.53$ \textcolor{black}{s}, the third platoon operates under the same lossy V2V channel but with the headway chosen as $0.6$ \textcolor{black}{s}, resulting in a string stable platoon. A headway of $0.6$ \textcolor{black}{s} is smaller than the minimum for an ACC platoon ($0.8$ \textcolor{black}{s}) and that for a lossy one-vehicle lookup platoon ($0.73$ \textcolor{black}{s}) \cite{vegamoor2019}, so there is no need to degrade the platooning mode.

\subsection{Higher-Fidelity Longitudinal Model}\label{sec:High_fidelity}
Since we are concerned about longitudinal string stability, it is sufficient to  capture the behavior of the drive-line and braking system of a vehicle, ignoring lateral dynamics. A variety of longitudinal models are available in literature depending on components of interest (engine/transmission/tires) and level of fidelity required \cite{LongModel1, LongModel2, LongModel3}. Many of them either require extensive data collection or privileged information from the vehicle/component manufacturer. \textcolor{black}{Open source simulators like SUMO \cite{SUMO_paper} ignore the actuation lags characteristic of real vehicles and past researchers in this area have often used simplified linear models for their validation \cite{AccianiObserver_Conf, JointCommDelay}}. Instead, we utilize an empirical vehicle model for validation, similar to \cite{vegamoorMS} and develop throttle/brake maps that relate pedal inputs and vehicle speed to acceleration generated. These signals are typically available directly on the onboard CAN bus of any drive-by-wire capable vehicle. In our case, an AutonomouStuff instrumented 2017 Lincoln MKZ hybrid car was used. Unlike in \cite{vegamoorMS}, there was no need to model the transmission \textcolor{black}{explicitly} since the MKZ hybrid car uses a continuously variable transmission.

\begin{figure}[htbp]
	\centering
	\includegraphics[width=9cm]{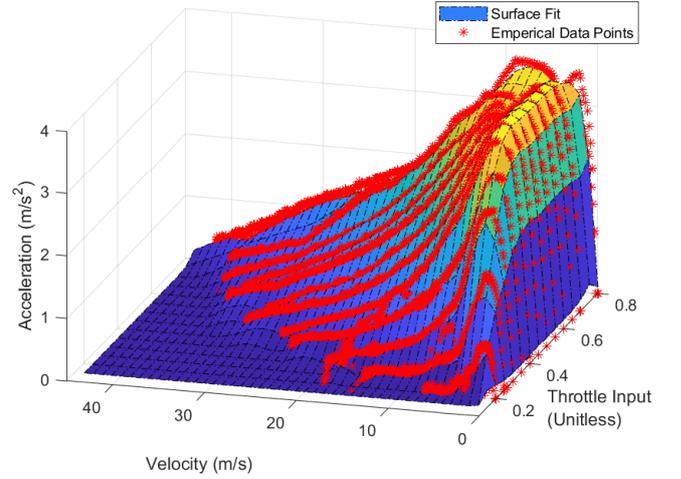}
	\caption{Throttle map of 2017 Lincoln MKZ}
	\label{fig:ThrottleMap}
\end{figure}

\begin{figure}[htbp]
	\centering
	\includegraphics[width=9 cm]{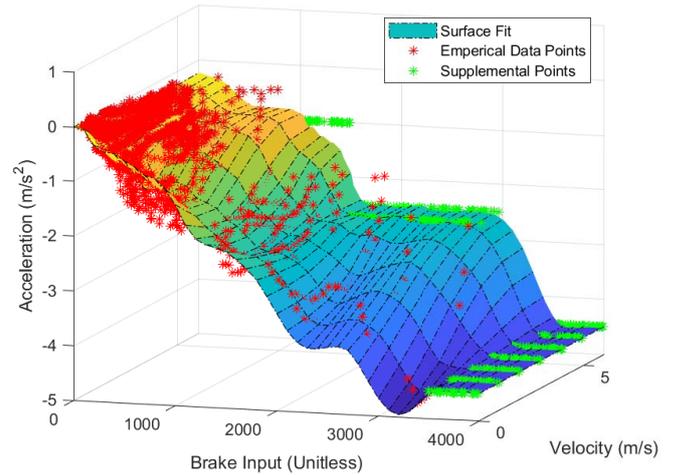}
	\caption{Brake map of 2017 Lincoln MKZ}
	\label{fig:BrakeMap}
\end{figure}

The throttle and brake maps are presented in Figs. \ref{fig:ThrottleMap} and \ref{fig:BrakeMap}. Data was collected by cycling through different combinations of pedal inputs and velocities. Supplemental points were added manually at the extremities of the brake map to saturate the deceleration estimates and for smoother interpolation. The surface fit was obtained using `gridfit' function in MATLAB. These maps, the vehicle model in Simulink are available in a Github repository \cite{githubCACC}.

\begin{figure}[htbp]

	\includegraphics[width=8cm]{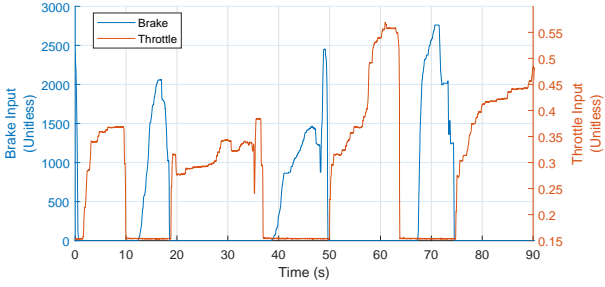}
	\caption{Brake and throttle inputs used for validation}
	\label{fig:MKZValid_Inputs}
\end{figure}
\begin{figure}[htbp]
	\centering
	\includegraphics[width=8.5cm]{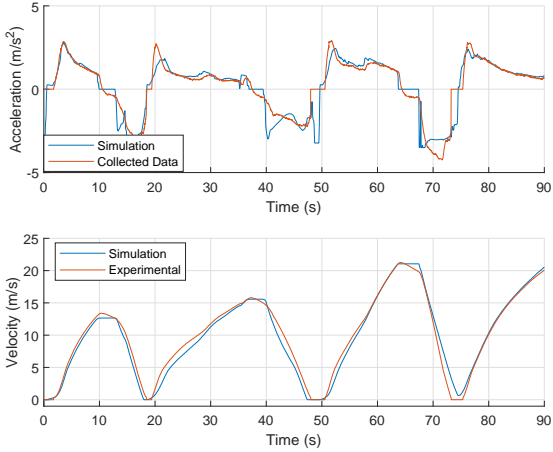}
	\caption{Model validation of longitudinal acceleration and velocity}
	\label{fig:MKZValid_VelAccel}
\end{figure}

To validate the model developed, a test run was performed on the real vehicle through manual driving. The throttle and brake inputs were recorded (as shown in Fig \ref{fig:MKZValid_Inputs}) and the same was supplied to the longitudinal model in simulation. The recorded acceleration and velocity of the real vehicle is compared with the output of the simulated vehicle in Fig. \ref{fig:MKZValid_VelAccel}.

As we can see, the developed model is able to capture the longitudinal dynamics of the real vehicle and predict the variables of interest (acceleration and velocity) with sufficient fidelity. Position of the vehicle is obtained through integration and is not as important for model validation as the platoon controllers only require relative position while they require absolute velocity and absolute acceleration. Next, we will use this newly validated model to corroborate the lossy CACC and CACC+ control schemes for a variety of time headway settings.

\subsection{CACC/CACC+ Simulations with Validated Car Model}\label{sec:CACC+HighFidSims}
\begin{figure}[htbp]
	\centering
	\includegraphics[width=9.5cm, height=7.5cm]{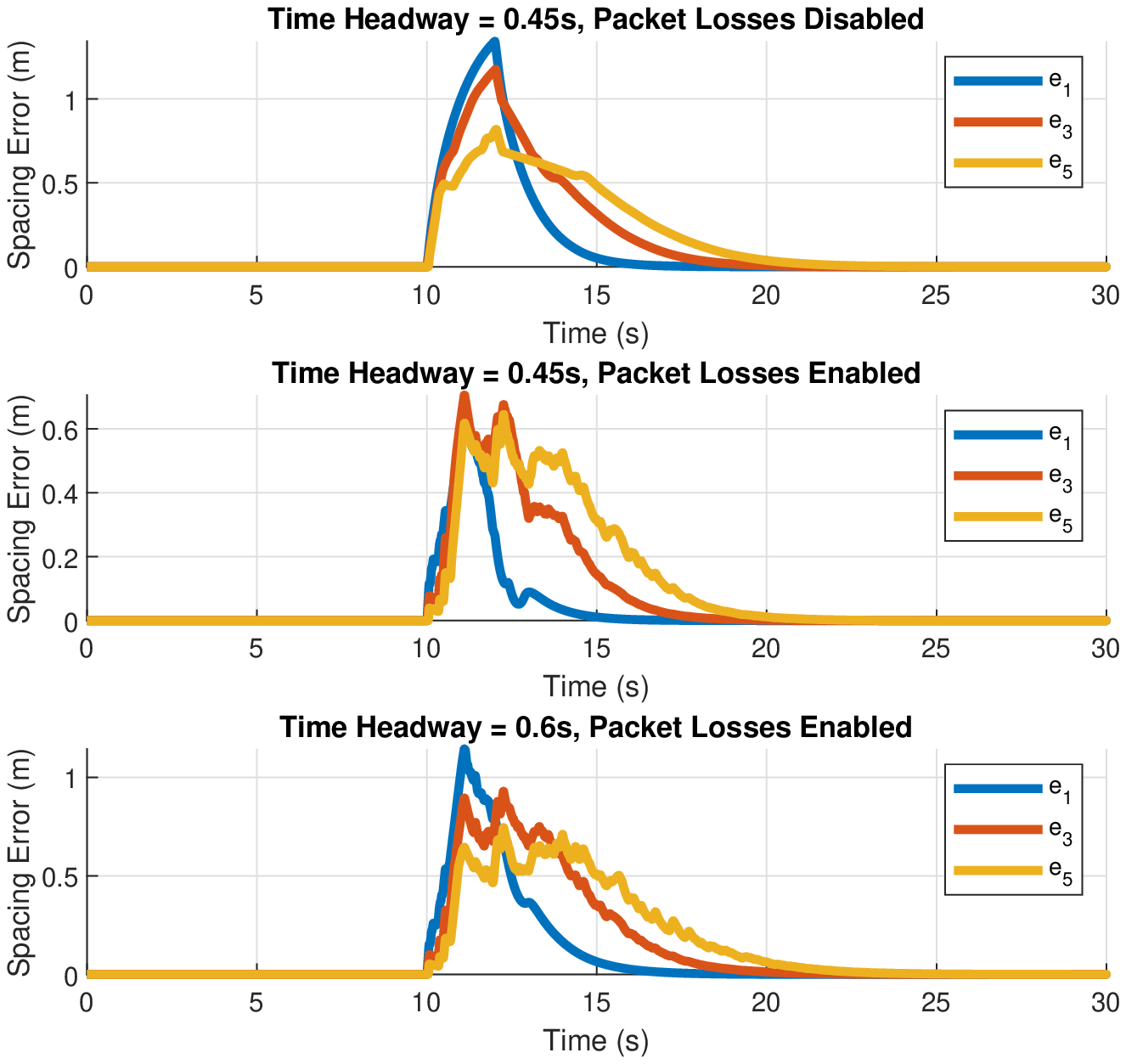}
	\caption{Spacing errors a CACC platoon with high fidelity model under different scenarios}
	\label{fig:MKZ1Veh}
\end{figure}

We use the same Gilbert burst channel parameters and the same lead vehicle maneuvers as in Section \ref{sec:LinearModelSims}. For lossy one vehicle lookup (CACC), the following controller gains were used: $(K_a,K_v, K_p) =(0.8,1.5,2)$. Actuation braking lag in the Lincoln MKZ was measured to be $0.37$ \textcolor{black}{s}, based on the deceleration step response on the real vehicle. This value was used for $\tau$ to calculate the minimum time headway. Three scenarios are presented in Fig. \ref{fig:MKZ1Veh} with a platoon of validated virtual vehicles: first without any losses and a time headway of $0.45$ \textcolor{black}{s}, then with losses enforced in the V2V link, and finally after increasing the time headway to $0.6$ \textcolor{black}{s}. 

For lossy one-vehicle lookup, the sufficient minimum condition for headway, from equation (\ref{eqn:PL_CACC_Limit}), is $0.538$ \textcolor{black}{s}. So as expected, an adjusted headway of $0.6$ \textcolor{black}{s} provides string stability with the spacing errors diminishing across the platoon, while a headway of $0.45$ \textcolor{black}{s} is unstable if the communication link is not ideal. There is no need to degrade the platoon to ACC mode (for which the sufficient condition on the minimum time headway is $2\tau = 0.74$ \textcolor{black}{s}). 

\begin{figure}[htbp]
	\centering
	\includegraphics[width=9.5cm, height=7.5cm]{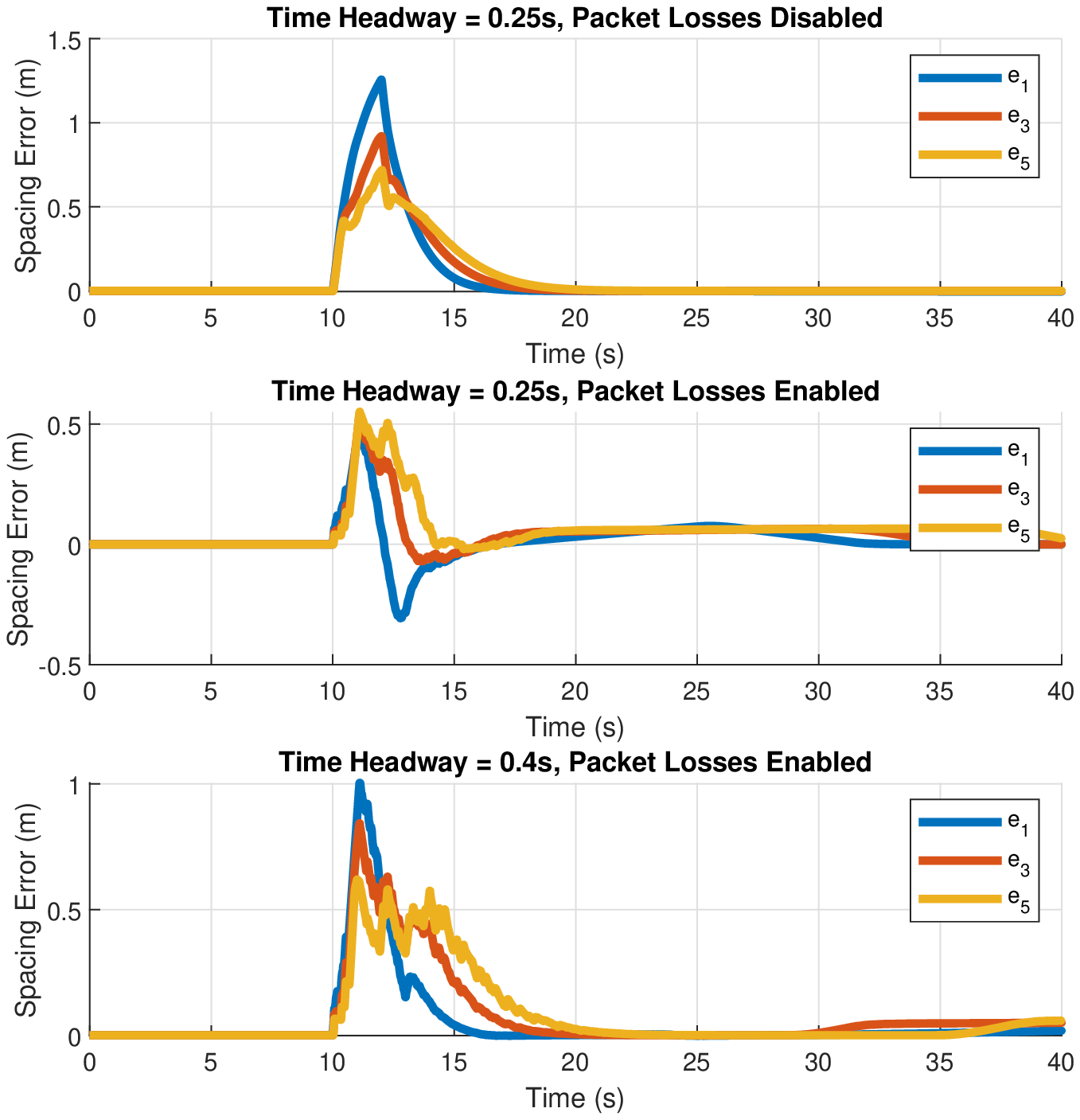}
	\caption{Spacing errors a CACC+ platoon with high fidelity model under different scenarios}
	\label{fig:MKZ2Veh}
\end{figure}

Similarly, three scenarios for a two vehicle (CACC+) scheme are presented in Fig. \ref{fig:MKZ2Veh}. The gains used were: $(K_a, K_v, K_p)= (0.75,2.5,1.5)$. Again, we observe that a time headway that would otherwise be stable under ideal V2V communication becomes unstable when packet losses are introduced. The minimum headway for the given value of $\gamma$ from equation (\ref{eqn:CACC+_hwLimit}) is $0.371$ \textcolor{black}{s} so picking a headway of $0.4$ \textcolor{black}{s} stabilizes the platoon, without the need to degrade to a CACC scheme.

\FloatBarrier

 \section{Conclusion}
In this work, we proposed a method to uniformly bound spacing errors for any vehicle in a \textcolor{black}{homgeneous} platoon, given the platoon leader's motion which is relevant from a safety perspective. Earlier results for a sufficient condition on the minimum string stable time headway for lossy one-vehicle lookup schemes were validated for burst noise channels. Furthermore, an approximate estimate of the same for a two-vehicle lookup scheme was also presented. Finally, the time headway constraints were corroborated using a high fidelity longitudinal model that was validated on a 2017 Lincoln MKZ hybrid car.

\color{black}
From a fuel-efficiency and traffic congestion standpoint, it is important that platoons are able to achieve the smallest inter-vehicular gap while maintaining safety, even in non-ideal environments like urban canyons where packet reception is poor. In such scenarios, this paper serves as a guideline for selecting the smallest stand still distance as well as the time headway - the two components that define the inter-vehicular gap in a platoon.

Furthermore, in this paper, we have only considered one and two-vehicle predecessor schemes. The approach from Section 
\ref{sec:lossyCACC+} maybe extended to multiple vehicle schemes and time headway limits similar to equation (\ref{eqn:CACCPlushwLimitMu}) may be derived with some involved algebra. That said, the limits derived in such manner for higher number of predecessors may not always be the smallest attainable time headway for the platoon. It is possible that as the complexity of the communication topology (number of predecessors) increases, certain combination of packet reception rates and controller gains may yield situations where it would be advisable to disable the link completely. That is, situations could arise where moving from an $(N)$ vehicle lookup to $(N-1)$ vehicle lookup scheme yields smaller achievable time headway.

\textcolor{black}{For future work, it would be interesting to explore conditions for stability when not only the communicated information, but also the on-board information is lost. For example, in situations with faulty radar or GPS sensors, the position and velocity information may not be available to the controller, requiring an adjustment to the time headway employed. Furthermore, we have ignored ride comfort in this paper and focused on platoon stability. Ensuring a smooth ride by minimizing jerk while guaranteeing string stability would also be a worthwhile pursuit.}

\color{black}
 Acknowledgment: This work was supported by US Department of Transportation (FHWA) through fellowship number 693JJ32045024 and Safety Through Disruption (SAFE-D) University Transportation Center, project $04-117$.
\bibliographystyle{IEEEtran}
\bibliography{Journal}
 \section{Appendix}
 \subsection*{Expected Value of Powers of Random Matrices} \label{subSec:App_RandomMatrix}
 \color{black}
 Suppose $A_0 \in \Re^{3 \times 3}$; for $i=1, 2, \ldots $, let  $b_{i} \in \Re^3$ and $e_1, e_2, e_3$ are the three column vectors of the $3 \times 3$ identity matrix. 
 For $i=1,2, \ldots, $ let $B_{i} = e_3 b_{i}^T$ so that we may define 
 $$A_L = \left[\begin{array}{ccc}
 A_0&0&0\\B_{1}&A_0&0\\
 0&B_{2}&A_0 \end{array} \right]. $$
 In the one-vehicle lookup (CACC) case, $A_0$ is deterministic and all vectors $b_i$ are independent of each other (due to independence of V2V links). Clearly,
 $$A_L^2 = \left[\begin{array}{ccc}
 A_0^2&0&0\\
 B_1A_0+A_0B_1 & A_0^2 &0 \\ B_2B_1 & B_2A_0 + A_0B_2 & A_0^2 \end{array} \right]. $$
Let
 $$A_L^k = \left[\begin{array}{ccc}
 A_0^k&0&0 \\ 
 A_{1,k} &A_0^k&0 \\
A_{3,k}&A_{2,k}&A_0^k 
 \end{array} \right],$$
 so that we can recursively write:
 \begin{eqnarray*}
 A_{1,k+1} &=& B_1A_0^k + A_0 A_{1,k} \\
 A_{2,k+1} &=& B_2A_0^k+A_0A_{2,k}\\
 A_{3,k+1} &=& B_2A_{1,k} + A_0 A_{3,k}. 
 \end{eqnarray*}
 Inductively, it is obvious that 
 \begin{itemize}
     \item $A_{1,k}$ is a linear function of $B_1$.
     \item $A_{2,k}$ is a linear function of $B_2$, and 
     \item $A_{3,k}$ is a bilinear function of $B_1$ and $B_2$. 
 \end{itemize}
 Since $b_1, b_2$ are linear functions of random variables with expectation ${\bar b}_1, {\bar b}_2$, we may conclude that  
 \begin{eqnarray*}
\mathbb{E}[A_{1,1}] &=& \mathbb{E}[B_1] = e_3 {\bar b}_1^T , \\ 
\mathbb{E}[A_{2,1}] &=& \mathbb{E}[B_2] = e_3{\bar b}_2^T,  \\
\mathbb{E}[A_{1,k+1}] &=& \mathbb{E}[B_1]A_0^k + A_0\mathbb{E}[A_{1,k}], \\
\mathbb{E}[A_{2,k+1}] &=& \mathbb{E}[B_2]A_0^k + A_0 \mathbb{E}[A_{2,k}], \\
\mathbb{E}[A_{3,k+1}] &=& \mathbb{E}[B_2]\mathbb{E}[A_{1,k}] + A_0 \mathbb{E}[A_{3,k}]. 
 \end{eqnarray*}
 Note that $A_{3,1}= 0$ and hence, $\mathbb{E}[A_{3,1}] = 0$. Consequently, we can conclude that 
 $$\mathbb{E}[A_L^{k+1}] = \mathbb{E}[A_L] \mathbb{E}[A_L^k], $$
 and hence, 
 $$\mathbb{E}[A_L^{k}] = (\mathbb{E}[A_L])^k. $$
\color{black}
\vspace{-0.5cm}
\section*{}
  \parpic{\includegraphics[width=1in,clip,keepaspectratio]{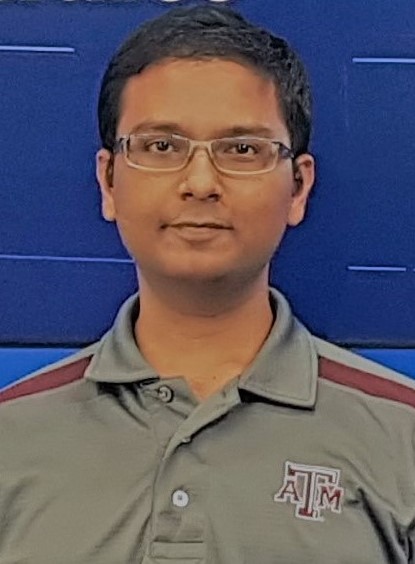}}
  \noindent {\textbf{Vamsi Vegamoor}} is currently pursuing his PhD with the department of Mechanical Engineering at Texas A\&M University, having received his MS from the department in 2018. His research focuses on vehicle spacing policies and sensor fusion for automated vehicles.
  \\
   \parpic{\includegraphics[width=1in,clip,keepaspectratio]{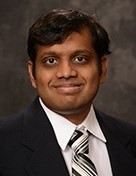}}
   
  \noindent {\textbf{Sivakumar Rathinam}} received his PhD from University of California at Berkeley in 2007. He is currently an Associate Professor with the Mechanical Engineering Department at Texas A\&M University. His research interests include motion planning and control of autonomous vehicles, collaborative decision making, combinatorial optimization, vision-based control, and air traffic control.
  \\
   \parpic{\includegraphics[width=1in,clip,keepaspectratio]{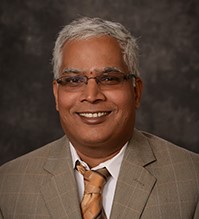}}
   
  \noindent {\textbf{Swaroop Darbha}} is a Professor in Mechanical Engineering at Texas A\&M University. He received his PhD from University of California at Berkeley in 1994. He is a fellow of ASME and IEEE for his contributions to Intelligent Transportation Systems. Dr. Darbha's research interests are on dynamics, control and diagnostics of connected and autonomous ground vehicles, routing of unmanned aerial vehicles, and decision-making under uncertainty.

\clearpage
\newpage
\onecolumn
\end{document}